# SHADES

## $^{22}$Ne(α,n)$^{25}$Mg reaction rate in the Gamow window

*David* Rapagnani[1], *Chemseddine* Ananna[1], *Antonino* Di Leva[1], *Gianluca* Imbriani[1], *Matthias* Junker[2], *Marco* Pignatari[3,4,5,6], and *Andreas* Best[1,*]

[1]Università di Napoli "Federico II" and INFN - Sezione di Napoli, 80126 Napoli, Italy
[2]INFN, Laboratori Nazionali del Gran Sasso (LNGS), 67100 Assergi, Italy
[3]Konkoly Observatory, Research Centre for Astronomy and Earth Sciences, Hungarian Academy of Sciences, Konkoly Thege Miklos ut 15-17, H-1121 Budapest, Hungary
[4]E.A. Milne Centre for Astrophysics, Department of Physics and Mathematics, University of Hull, HU6 7RX, United Kingdom
[5]Joint Institute for Nuclear Astrophysics, Center for the Evolution of the Elements, Michigan State University, 640 South Shaw Lane, East Lansing, MI 48824, USA
[6]NuGrid Collaboration, nugridstars.org

**Abstract.** Neutron capture reactions are the main contributors to the synthesis of the heavy elements through the s-process. Together with $^{13}$C(α,n)$^{16}$O, which has recently been measured by the LUNA collaboration in an energy region inside the Gamow peak, $^{22}$Ne(α,n)$^{25}$Mg is the other main neutron source in stars. Its cross section is mostly unknown in the relevant stellar energy (450 keV < $E_{cm}$ < 750 keV), where only upper limits from direct experiments and highly uncertain estimates from indirect sources exist. The ERC project SHADES (UniNa/INFN) aims to provide for the first time direct cross section data in this region and to reduce the uncertainties of higher energy resonance parameters. High sensitivity measurements will be performed with the new LUNA-MV accelerator at the INFN-LNGS laboratory in Italy: the energy sensitivity of the SHADES hybrid neutron detector, together with the low background environment of the LNGS and the high beam current of the new accelerator promises to improve the sensitivity by over 2 orders of magnitude over the state of the art, allowing to finally probe the unexplored low-energy cross section. Here we present an overview of the project and first results on the setup characterization.

## 1 Introduction

In massive stars (M>8M$_\odot$) $^{22}$Ne(α,n)$^{25}$Mg is the main neutron source for the synthesis of the isotopes made by the slow neutron capture process (s-process) in the mass region A 6090, between iron and the first neutron magic peak (N=50) [1]; together with $^{13}$C(α,n)$^{16}$O, which has recently measured with high precision by LUNA collaboration ([2] and references therein) it is also responsible for the synthesis of s-process elements with mass $A \sim 90 - 209$ in Asymptotic Giant Branch (AGB) stars [3–6]. Despite of the many efforts in measuring $^{22}$Ne(α,n)$^{25}$Mg during the last fifty years (e.g. [7–11]), present uncertainty on reaction rate

---

*

e-mail: andreas.best@na.infn.it

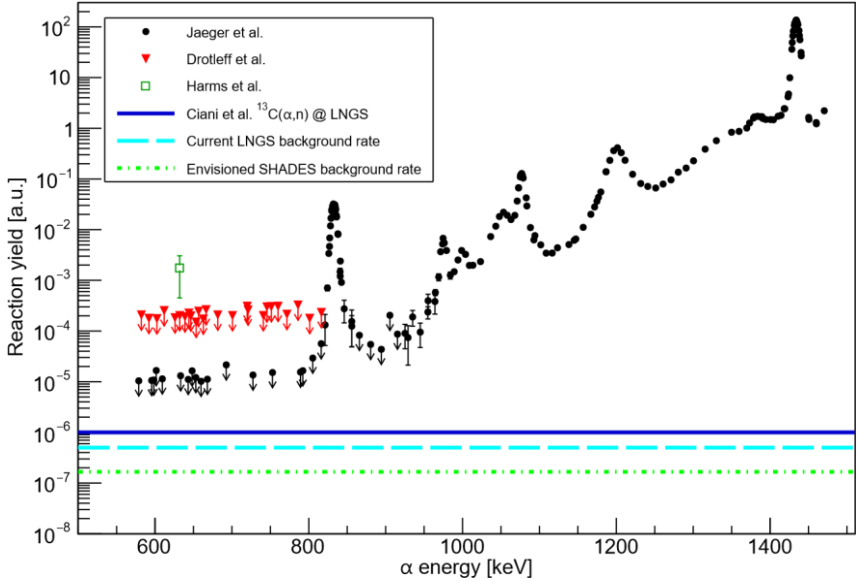

Figure 1. Reaction yield data from Jaeger et al. ([8], black circles), Drotleff et al. ([9], red triangles) and Harms et al. ([10], green rectangle). The background level estimated for the $^{13}$C($\alpha$,n)$^{16}$O LUNA measurement campaign [2] and the one envisioned for SHADES are also shown. See the text for more details.

estimate reaches a value of about 500% at temperature around 0.2-0.3 GK, the one of interest for helium burning. In addition, only upper limit values are known at the moment for the cross section, whereas the knowledge of its actual value is necessary for stellar models. For this the European Research Council founded project SHADES (Scintillator-He3 Array for Deep-underground Experiments on the S-process) is developing an innovative low background measurement campaign which will be described in the following sections.

## 2 SHADES

The challenging experimental condition limited the sensitivity of measurements performed so far. As it is clear from Figure 1, only upper limits for the reaction cross section at energies below $E_\alpha \lesssim 800$ keV are currently known. Since the efforts of Jeager et al. [8], which reduced by one order of magnitude the upper limit, no progress was made on $^{22}$Ne($\alpha$,n)$^{25}$Mg while a greater sensitivity is still required to measure actual values of its cross section. The SHADES project goal is to improve the experimental sensitivity by at least two order of magnitude. This increase in sensitivity will be obtained by (1) performing the reaction cross section measurement underground at Gran Sasso National Laboratory (LNGS) of the Italian Institute of Nuclear Physics (INFN), (2) employing the new high current LUNA-MV ion beam accelerator, (3) with the use of an innovative detection array which grants a selective discrimination of reaction neutrons and (4) with an high purity $^{22}$Ne gas target system. This efforts will allow to perform (5) precise astrophysical estimates of s-process influenced elemental abundance, solving a longstanding issue on this important astrophysical process.

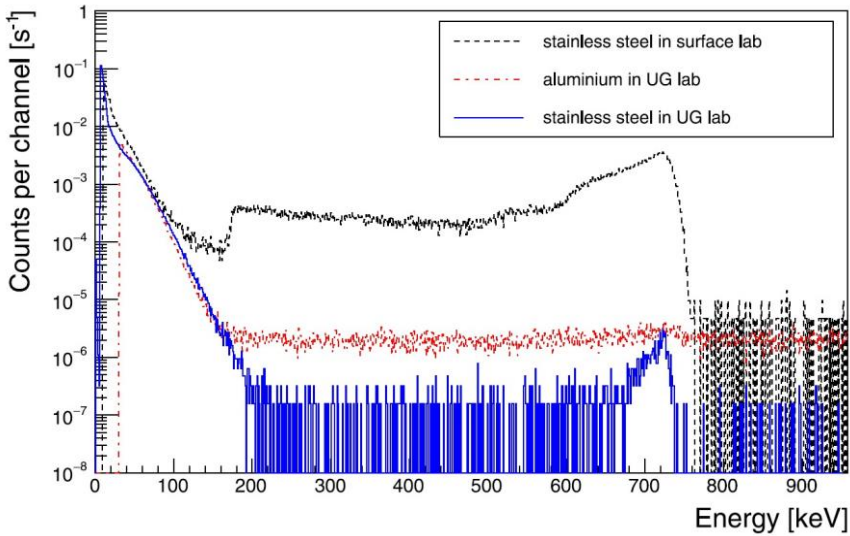

Figure 2. The spectra, as presented in [12], for surface measurements with stainless steel He3 proportional counters (black dots) and underground measurements with both aluminum (red dots) and stainless-steel (blue line) He3 proportional counters. See also [13] for more details.

### 2.1 Underground Laboratory

Being deep underground, LNGS offer a unique environment for low neutron background measurements. In this respect, Figure 2 shows the spectra of stainless steel and aluminum He3 proportional counters tested on surface laboratory and at LNGS: the natural neutron flux suppression is so high underground that the intrinsic activity of aluminum becomes the main source of background.

### 2.2 LUNA-MV

The new LUNA-MV accelerator, currently under installation at LNGS, is a 3.5 MV Singletron Accelerator [14] with an ECR source capable of deliver up to 500 μA of high energy resolved alpha particles beams. It has also an extremely high stability over time, a feature desirable since the long measurement runs expected for SHADES.

### 2.3 Array

The SHADES array is composed by 12 EJ-309 liquid scintillators and 18 He3 counters held by a borated polyethylene support which acts also as a shield for environmental neutrons (Figure 3). A reaction neutron will be recorded only after being moderation inside the scintillators and capture in the He3 proportional counters. Together with a Pulse Shape Discrimination analysis [15], this procedure allows a further background suppression.

### 2.4 Gas Target

In order to reach the sensitivity necessary to measure $^{22}$Ne($\alpha$,n)$^{25}$Mg at the lower energies, beam induced background events must be kept at minimum. Carbon and boron contamination, due to their higher ($\alpha$,n) cross section, can be a formidable source of background neutrons. For this, a new recirculating high purity 30 keV thick Ne22 extended gas target was designed and is presently under commissioning in the Nuclear Astrophysics Laboratory of University of Naples "Federico II". A combination of high flow and high vacuum pumps in different pumping stage and differential configuration allow to reach the high vacuum conditions of LUNA-MV of $\lesssim 10^{-7}$ mbar from target region pressure of about 5 mbar in a few centimeters of beam line. Indeed, to optimize detection efficiency, the gas target dimensions has been kept as low as possible. Carbon-free pumps, gaskets and sealings and a gas purifier were also adopted to reduce carbon contaminant from vacuum elements degassing and air leaks. A detector at 165° with respect the ion beam direction will allow to measure back-scattered $\alpha$-particles for ion beam current monitoring. Several pressure and temperature gauges will serve as stability monitor for gas target during the long reaction cross section measurement runs.

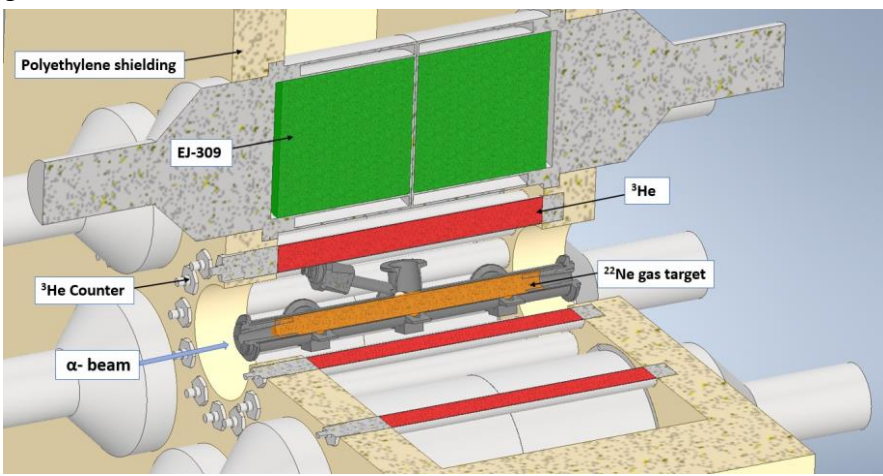

Figure 3. Cut-view of the SHADES detectors array and gas target setup.

### 2.5 Astrophysical Evaluations

After an R-Matrix analysis of the cross section data, the astrophysical impact of SHADES experimental outcomes will be evaluated using the NuGrid codes and their stellar library for a complete set of stellar masses and metallicities [16–19].

## 3 Outlook

The ERC-funded project "Scintillator-He3 Array for Deep-underground Experiments on the S-process" (SHADES) aims to measure $^{22}$Ne($\alpha$,n)$^{25}$Mg down to neutron threshold energy with an innovative detection array and a low-background environment granted by the LNGS underground laboratories and an extremely pure gas target system. The detectors array and gas target design is already complete, while the array characterization is ongoing on surface laboratory in the University of Naples "Federico II" and underground at LNGS. The gas target is currently under installation and commissioning and will enter soon in the characterization phase through Ion Beam Analysis measurements to be performed at the Tandem Accelerator laboratory of the University of Campania in Caserta during winter 2021. The preparation of underground measurements is expected for Spring-Winter 2022 while the $^{22}$Ne($\alpha$,n)$^{25}$Mg rate measurements is planned for the whole 2023. Finally R-Matrix analysis and nucleosynthesys calculations will be performed during 2024.

### Acknowledgement

The authors acknowledge funding from the European Research Council (ERC-StG 2019 #852016). MP acknowledges support to NuGrid from STFC (through the University of Hull's Consolidated Grant ST/R000840/1), and access to viper, the University of Hull HPC Facility. MP also Acknowledges the National Science Foundation (NSF, USA) under grant No. PHY1430152 (JINA Center for the Evolution of the Elements), the "Lendulet-2014" Program of the Hungarian Academy of Sciences (Hungary), the ERC Consolidator Grant funding scheme (Project RADIOSTAR, G.A. n. 724560, Hungary), the ChETEC COST Action (CA16117), supported by the European Cooperation in Science and Technology, the ChETEC-INFRA project funded from the European Union's Horizon 2020 research and innovation programme (grant agreement No 101008324), and the IReNA network supported by NSF AccelNet.